\documentstyle[prl,aps]{revtex}
\begin{document}

\begin{center}
{\bf Reply to the comment by Jacobs and Thorpe}
\end{center}

We welcome the interesting comment by Jacobs and Thorpe (JT)[1], and
 are gratified that first-order rigidity is confirmed
in their random bond model.   
We have worked with them to look again at the rigidity of
Cayley trees, with the conclusion that the $p_c$ on trees
discussed our letter[2] is analogous to a spinodal point
in thermodynamic transitions. In fact
we have developed a new constraint equation, which when
 combined with our Cayley tree theory[3], leads to results
 which are numerically equivalent to the 
random bond model results presented in their comment. 
JT also argue that in two dimensions the ``transition'' is second
order.  We agree.  
 There is a non-trivial diverging correlation
length, so the ``transition'' is second order[2].  However, we raise
the possibility in[2] that the ``order parameter'' has a first-order
jump or a very small exponent.  JT imply that first order rigidity is
pathalogical. We {\it disagree}.    
We think that first-order rigidity is already well known experimentally.
A common example is the collapse of a granular solid.  Such
a solid may be supported by internally sintered grains or
by an external boundary.  As these constraints are removed, the
solid may eventually collapse in a first-order fashion
(e.g. a landslide).  This example is however a ``directed''
rigidity problem in that the central force constraints
are  primarily in compression. In detail our interpretation of the current
data on trees and on ``generic'' triangular lattices is as follows:

 (i) Within {\it Cayley tree models}, $P_{\infty}$[2-4] 
and $f'$[1] both have jump discontinuities.  The transition is 
then {\it first order}.  The analysis of the random-bond problem by JT is for 
free boundaries, while our Cayley tree analysis[2,4] was 
for a rigid boundary.   Motivated by the JT comment,
we have analysed the tree model for a variety of boundaries.
In addition we have found an expression for
$f' = (-z/2g)(1-T_{\infty}^2)$ on trees[3]. 
The solution to the Cayley-tree equations, for
$z=6$, $g=2$ and a rigid boundary gives $p_c=0.605$[3].
However if we introduce a consistency equation[3] which removes
boundary constraints, we find $p_c=0.655$ (same as JT's
random-bond model).  In fact, in this case, our results for $f'$ are
equivalent to theirs. 
 In the regime $0.605<p<0.655$, the solutions to the 
Cayley tree equations can be interpretted as
being {\it metastable}.  It is
an open question as to whether this ``spinodal'' regime is
accessable in granular solids.

(ii)  On the {\it triangular lattice}, there is a 
diverging correlation length and a non-trivial 
correlation-length exponent[2,5,6]. 
Thus, we (and JT) have clearly demonstrated that 
the rigidity transition on the triangular lattice
is {\it second order}[7], though whether it is 
``straightforward'' or ``conventional''[1]  is a
semantic issue[8].
  On triangular lattices, 
we also both agree that the 
backbone goes to zero continuously with 
($\beta'=0.25\pm 0.03$[2,6], $\beta'=0.24 \pm 0.04$[5]).
There is however a real difference in our interpretations of the data for  
the {\it infinite-cluster probability},
$P_{\infty}\sim (p-p_c)^{\beta}$.
Taken alone, we think that the
current numerical data on the triangular lattice
 is consistent with any $\beta < 0.175$.
In particular, in our letter[2], we argued for a first
order jump at $p_c$ ($\beta \sim  0$) based
primarily on the analysis of the ``difference'' or 
dangling ends.  In contrast
JT [5] analyse $P_{\infty}$ itself
and find ($\beta \sim 0.175\pm 0.02$).  
There are very large finite-size effects
in this problem, so that $L\sim 100,000$ is needed 
in order to find $\beta$ precisely in two dimensions, both
for the ``difference''(see[1]) and for the infinite cluster itself. 
With careful use of memory this appears achievable in the next couple of years,
using refinements of the current algorithms.

We thank Don Jacobs and Mike Thorpe 
for stimulating discussions.  PMD thanks
 the DOE for support under
contract DE-FG02-90ER45418 and CM acknowledges support 
from the Conselho Nacional de Pesquisa CNPq, Brazil.\\

\noindent --- P.M. Duxbury,\\
Department of Physics and Astonomy, \\
Michigan State University, 
East Lansing, 48824.\\
--- C. Moukarzel,\\
Instituto de F\'isica, Universidade Federal Fluminense,\\ 
CEP 24210-340, Niteroi RJ, Brazil\\
--- P.L. Leath,\\
Department of Physics and Astronomy,\\ 
Rutgers University, 
Piscataway, New Jersey 08855-849\\

\noindent  [1] D. Jacobs and M.F. Thorpe, preceeding comment\\
\noindent [2] C. Moukarzel, P.M. Duxbury and P.L. Leath.
Phys. Rev. Lett. {\bf 78}, 1480 (1997)\\
\noindent [3] P.M. Duxbury, D. Jacobs, M.F. Thorpe and C. Moukarzel preprint\\
\noindent [4] C. Moukarzel, P.M. Duxbury and P.L. Leath,
Phys. Rev. {\bf E55}, 5800 (1997)\\
\noindent [5] D. Jacobs and M.F. Thorpe, Phys. Rev. Lett. {\bf 75},
4051 (1995); Phys. Rev. {\bf E53}, 3682 (1996)\\
\noindent [6] C Moukarzel and P.M. Duxbury, 
Phys. Rev. Lett. {\bf 75}, 4055 (1995)\\
\noindent [7] It is also well established that 
the rigidity transition in two dimensions 
($\nu = 1.16 \pm 0.03$)[2,6], $\nu = 1.21 \pm 0.06$[5])
 is in a different universality class 
than connectivity percolation ($\nu = 4/3$)\\
\noindent [8] A jump discontinuity in an order
parameter does occur in some  ``conventional'' second order
transitions - e.g. the helicity modulus of the 2-d xy model.
The infinite cluster probability is like an order parameter of
percolation problems.\\

\end{document}